\documentclass[preprint,twocolumn]{aastex631}

\usepackage{float}

\newcommand{\objname}{(248370)~2005~QN$_{173}$}
\newcommand{\totalthumbs}{eighty-one}
\newcommand{\totalobservations}{thirty-one}

\shorttitle{Active Asteroid (248370)~2005~QN$_{173}$ Recurrent Activity}
\shortauthors{Chandler, Trujillo \& Hsieh}

\usepackage{graphicx}
\usepackage[percent]{overpic} 

\begin{document}

\title{Recurrent Activity from Active Asteroid (248370)~2005~QN$_{173}$: A Main-belt Comet}

\correspondingauthor{Colin Orion Chandler}
\email{orion@nau.edu}

\author[0000-0001-7335-1715]{Colin Orion Chandler}
\affiliation{Department of Astronomy and Planetary Science, Northern Arizona University, PO Box 6010, Flagstaff, AZ 86011, USA}

\author[0000-0001-9859-0894]{Chadwick A. Trujillo}
\affiliation{Department of Astronomy and Planetary Science, Northern Arizona University, PO Box 6010, Flagstaff, AZ 86011, USA}

\author[0000-0001-7225-9271]{Henry H. Hsieh}
\affiliation{Planetary Science Institute, 1700 East Fort Lowell Rd., Suite 106, Tucson, AZ 85719, USA}
\affiliation{Institute of Astronomy and Astrophysics, Academia Sinica, P.O.\ Box 23-141, Taipei 10617, Taiwan}


\begin{abstract}
\label{Abstract}
We present archival observations of main-belt asteroid \objname{} (also designated 433P) that demonstrate this recently discovered active asteroid (a body with a dynamically asteroidal orbit displaying a tail or coma) has had at least one additional apparition of activity near perihelion during a prior orbit. We discovered evidence of this second activity epoch in an image captured 2016 July 22 with the DECam on the 4~m Blanco telescope at the Cerro Tololo Inter-American Observatory in Chile. As of this writing, \objname{} is just the 8th active asteroid demonstrated to undergo recurrent activity near perihelion. Our analyses demonstrate \objname{} is likely a member of the active asteroid subset known as main-belt comets, a group of objects that orbit in the main asteroid belt that exhibit activity that is specifically driven by sublimation. We implement an activity detection technique, \textit{wedge photometry}, that has the potential to detect tails in images of solar system objects and quantify their agreement with computed antisolar and antimotion vectors normally associated with observed tail directions. We present a catalog and an image gallery of archival observations. The object will soon become unobservable as it passes behind the Sun as seen from Earth, and when it again becomes visible (late 2022) it will be farther than 3~au from the Sun. Our findings suggest \objname{} is most active interior to 2.7~au (0.3~au from perihelion), so we encourage the community to observe and study this special object before 2021 December.
\end{abstract}

\keywords{minor planets, asteroids: individual ((248370) 2005 QN173), comets: individual (433P)}

\section{Introduction}
\label{introduction}

\begin{figure*}
    \centering
    \includegraphics[width=1.0\linewidth]{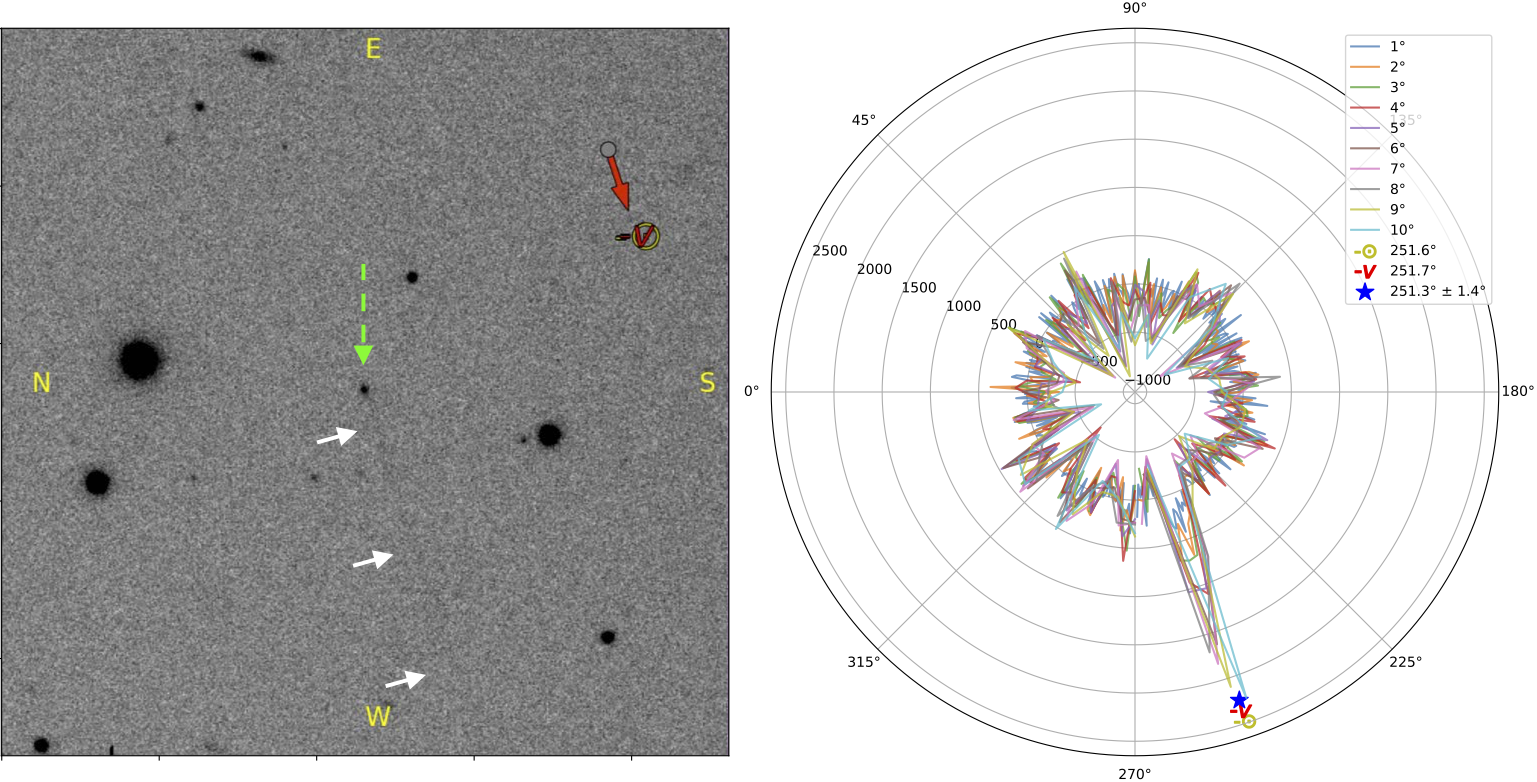}
    \caption{The 126\arcsec$\times$126\arcsec thumbnail image (left) shows \objname{} (green dashed arrow) at center with a tail (white arrows) oriented towards 5 o'clock. This 89~s $z$-band exposure captured with the DECam is the only image in which we could unambiguously identify activity. We conducted wedge photometry (right) that shows the tail orientation is $251\fdg3\pm1\fdg4$ (blue star), in close agreement with the $251\fdg6$ antisolar angle (yellow $\odot$) and the $251\fdg7$ antimotion vector (red $v$) as computed by JPL Horizons. The plot shows counts radially outward from the the object center at (0,0).}
    \label{fig:wedgephot}
\end{figure*}

Active asteroids are objects that are dynamically asteroidal but that display comet-like activity such as a tail or coma \citep{hsiehActiveAsteroidsMystery2006}. Activity may be caused by mechanisms unrelated to volatiles (e.g., impact, rotational disruption) or by sublimation as is typically the case with comets. Sublimation driven active objects provide key insights into the present-day volatile distribution in our solar system, as well as clues about the origins of those volatiles and how they arrived on Earth \citep{hsiehPopulationCometsMain2006}. These objects have been persistently difficult to study because of the small numbers detected to date: fewer than 30 active asteroids, of which roughly half are thought to exhibit sublimation driven activity; see \citealt{chandlerSAFARISearchingAsteroids2018} for a summary.

When the aforementioned sublimation driven activity is connected with a main-belt asteroid, the object is classified as a main-belt comet (MBC). MBCs are often characterized by activity near perihelion and the absence of activity elsewhere in the orbit \citep{hsiehMainbeltCometsPanSTARRS12015,agarwalBinaryMainbeltComet2017,hsieh2016ReactivationsMainbelt2018}, suggesting that the primary activity mechanism is sublimation of volatiles such as water ice \citep{snodgrassMainBeltComets2017}.

By contrast, stochastic events like impacts may result in comet-like activity but, in such cases, the appearance of activity is expected to cease once the material dissipates. Roughly 60\% of known active asteroids have been observed to display activity during only a single apparition \citep{chandlerSAFARISearchingAsteroids2018}.

Asteroid (7968), now comet 133P/Elst-Pizarro, was the first active main-belt asteroid to be discovered. While it was unclear at the time whether the activity was sublimation driven \citep{boehnhardt1996_133p,boehnhardt1998_133p} or due to a one-time event \citep{tothImpactgeneratedActivityPeriod2000}, subsequent apparitions showing activity indicated sublimation was the cause \citep{hsiehStrangeCase133P2004,hsiehReturnActivityMainbelt2010}. This example illustrates the importance of detecting additional activity epochs.

Asteroid \objname{} is a 3.2$\pm$0.4~km diameter \citep{hsiehPhysicalCharacterizationMainBelt2021} outer main-belt asteroid ($a$=3.075~au, $e$=0.226, $i$=$0\fdg067$) that has a 5.37~yr orbit that ranges from a perihelion distance of $q$=2.374~au to an aphelion distance of $Q$=3.761~au. The object first drew particular attention when it was reported as active on 2021 July 9 \citep{fitzsimmons2483702005QN1732021}. Subsequently, Zwicky Transient Facility (ZTF) data were used to help constrain the activity onset to between 2020 July 10 and 2021 June 11 \citep{kelley2483702005QN2021}. 

We set out to locate archival astronomical images of \objname{} in order to characterize prior activity. We made use of solar system object thumbnails (small image cutouts like Figure \ref{fig:wedgephot}) derived from publicly available archival data. We previously demonstrated how our data sources, such as the Dark Energy Camera (DECam), are well suited to discovering and characterizing active objects \citep{chandlerSAFARISearchingAsteroids2018,chandlerSixYearsSustained2019,chandlerCometaryActivityDiscovered2020a}.

Here we report activity of \objname{} on 2016 July 22 \citep{chandler2483702005QN2021}, an apparition prior to the 2021 outburst. We describe the process by which the activity was identified and examine the implications of this discovery.


\section{Second Activity Epoch}
\label{sec:secondActivityEpoch}

In order to find an additional activity epoch for \objname{}, we searched, assessed, and analyzed publicly available archival image data, building upon the methods of \cite{chandlerSAFARISearchingAsteroids2018,chandlerSixYearsSustained2019,chandlerCometaryActivityDiscovered2020a}. 

\subsection{Data Acquisition}
\label{subsec:datamining}

To locate archival images of \objname{}, we queried our own database of publicly available observation metadata \citep[see][]{chandlerSAFARISearchingAsteroids2018}. This database, which updates daily, includes observing details such as sky coordinates, exposure time, and filter selection. Additionally, we searched Palomar Transient Factory (PTF) and ZTF data through 2021 August 31 through online search tools (listed in Appendix \ref{sec:equipQuickRef}) as well as a ZTF Alert Stream search and retrieval tool we created for this purpose. All instruments and data sources we made use of are listed in Appendix \ref{sec:equipQuickRef}, and we note that some data were found or retrieved via more than one pathway.

\begin{table*}
\centering
\footnotesize
    \caption{\objname{} Observations}
    \begin{tabular}{cclcrcccrrccl}
        Image\footnote{Label in image gallery figures.} & Obs. Date\footnote{UT observing date in year-month-day format.} & Source      & N\footnote{Number of images.} & Exp. [s]\footnote{Exposure time for each image.} & Filter & $V$\footnote{Apparent $V$-band magnitude (Horizons).} & r [au]\footnote{Heliocentric distance.} & STO [$\degr$]\footnote{Sun--target--observer angle.}  & $\nu$ [$\degr$]\footnote{True anomaly.} & \%$_{Q\rightarrow q}$\footnote{Percentage to perihelion $q$ from aphelion $Q$.} & Act?\footnote{Activity observed.} & Archive\\
        \hline
a       & 2004-07-08               & MegaPrime   & 3                       & 180                       & \textit{i     } & 20.7                      & 2.74                  & 17.6                    & 287.5                 & 73\%                        & N    & CADC,*      \\
b       & 2005-06-08               & SuprimeCam  & 3                       & 60                        & \textit{W-J-VR} & 21.0                      & 2.42                  & 23.8                    & 18.6                  & 96\%                        & N    & CADC,SMOKA  \\
c       & 2010-06-14               & Pan-STARRS1 & 2                       & 30                        & \textit{z     } & 20.2                      & 2.42                  & 21.1                    & 339.2                 & 96\%                        & N    & CADC        \\
d       & 2010-08-02               & Pan-STARRS1 & 1                       & 45                        & \textit{i     } & 19.0                      & 2.39                  & 3.9                     & 353.5                 & 99\%                        & N    & CADC        \\
e       & 2010-08-05               & Pan-STARRS1 & 1                       & 40                        & \textit{r     } & 18.9                      & 2.39                  & 2.5                     & 355.4                 & 99\%                        & N    & CADC        \\
f       & 2010-08-06               & Pan-STARRS1 & 1                       & 43                        & \textit{g     } & 18.8                      & 2.39                  & 2.0                     & 354.7                 & 99\%                        & N    & CADC        \\
g       & 2010-08-28               & PTF         & 2                       & 60                        & \textit{r     } & 19.2                      & 2.39                  & 8.4                     & 1.2                   & 99\%                        & N    & IRSA/PTF    \\
h       & 2010-08-31               & Pan-STARRS1 & 2                       & 45                        & \textit{i     } & 19.3                      & 2.39                  & 9.7                     & 2.1                   & 99\%                        & N    & CADC        \\
i       & 2010-09-01               & PTF         & 2                       & 60                        & \textit{r     } & 19.3                      & 2.39                  & 10.1                    & 2.3                   & 99\%                        & N    & IRSA/PTF    \\
j       & 2010-09-06               & Pan-STARRS1 & 2                       & 43                        & \textit{g     } & 19.5                      & 2.39                  & 12.2                    & 3.8                   & 99\%                        & N    & CADC        \\
k       & 2010-09-12               & Pan-STARRS1 & 2,2 & 43,40 & \textit{g,r   } & 19.6                      & 2.40                  & 14.5                    & 5.6                   & 98\%                        & N    & CADC        \\
l       & 2010-09-15               & PTF         & 2                       & 60                        & \textit{r     } & 19.7                      & 2.39                  & 15.5                    & 6.5                   & 99\%                        & N    & IRSA/PTF    \\
m       & 2010-10-30               & Pan-STARRS1 & 2                       & 30                        & \textit{z     } & 20.6                      & 2.41                  & 23.8                    & 19.6                  & 97\%                        & N    & CADC        \\
n       & 2011-07-14               & Pan-STARRS1 & 1                       & 40                        & \textit{r     } & 21.8                      & 2.86                  & 15.8                    & 84.2                  & 65\%                        & N    & CADC        \\
o       & 2011-11-24               & Pan-STARRS1 & 2,2 & 43,40 & \textit{g,r   } & 20.6                      & 3.15                  & 4.0                     & 109.3                 & 44\%                        & N    & CADC        \\
p       & 2011-11-30               & Pan-STARRS1 & 2                       & 45                        & \textit{i     } & 20.4                      & 3.16                  & 1.8                     & 110.3                 & 43\%                        & N    & CADC        \\
q       & 2011-12-01               & Pan-STARRS1 & 2                       & 43                        & \textit{g     } & 20.4                      & 3.16                  & 1.4                     & 110.5                 & 43\%                        & N    & CADC        \\
r       & 2014-03-01               & OmegaCam    & 5                       & 360                       & \textit{r     } & 21.4                      & 3.53                  & 7.0                     & 218.2                 & 17\%                        & N    & CADC,ESO    \\
s       & 2016-07-22               & DECam       & 1                       & 89                        & \textit{z     } & 21.2                      & 2.59                  & 22.7                    & 56.5                  & 84\%                        & Y    & CADC,*      \\
t       & 2019-07-03               & DECam       & 9                       & 40                        & \textit{VR    } & 22.6                      & 3.55                  & 14.6                    & 216.9                 & 15\%                        & N    & CADC,*      \\
u       & 2020-02-04               & DECam       & 1                       & 38                        & \textit{r     } & 21.6                      & 3.16                  & 18.1                    & 248.9                 & 43\%                        & N    & CADC,*      \\
v       & 2020-02-10               & DECam       & 1                       & 199                       & \textit{z     } & 21.8                      & 3.15                  & 18.2                    & 249.9                 & 44\%                        & N    & CADC,*      \\
w       & 2020-04-25               & ZTF         & 1,1 & 30,30 & \textit{g,r   } & 20.0                      & 2.99                  & 4.4                     & 263.3                 & 55\%                        & N    & IRSA/ZTF    \\
x       & 2020-05-18               & ZTF         & 1,1 & 30,30 & \textit{g,r   } & 19.9                      & 2.93                  & 4.7                     & 267.7                 & 60\%                        & N    & IRSA/ZTF    \\
x       & 2020-05-27               & ZTF         & 1,1 & 30,30 & \textit{g,r   } & 20.1                      & 2.91                  & 8.4                     & 269.5                 & 61\%                        & N    & IRSA/ZTF    \\
x       & 2020-06-11               & ZTF         & 1,3 & 30,30 & \textit{g,r   } & 20.3                      & 2.88                  & 13.3                    & 272.5                 & 63\%                        & N    & IRSA/ZTF    \\
x       & 2020-06-14               & ZTF         & 1,1 & 30,30 & \textit{g,r   } & 20.4                      & 2.87                  & 14.1                    & 273.1                 & 64\%                        & N    & IRSA/ZTF    \\
x       & 2020-06-17               & ZTF         & 2,2 & 30,30 & \textit{g,r   } & 20.4                      & 2.87                  & 14.9                    & 273.8                 & 64\%                        & N    & IRSA/ZTF    \\
x       & 2020-06-20               & ZTF         & 2,2 & 30,30 & \textit{g,r   } & 20.5                      & 2.86                  & 15.7                    & 274.4                 & 65\%                        & N    & IRSA/ZTF    \\
x       & 2020-06-23               & ZTF         & 2,1 & 30,30 & \textit{g,r   } & 20.5                      & 2.85                  & 16.5                    & 275.0                 & 65\%                        & N    & IRSA/ZTF    \\
x       & 2020-06-26               & ZTF         & 4,1 & 30,30 & \textit{g,r   } & 20.6                      & 2.85                  & 17.2                    & 275.6                 & 65\%                        & N    & IRSA/ZTF\\
\hline
    \end{tabular}
    \footnotesize
    \raggedright
    Note \textit{W-J-VR} is a single wide-band filter. See Appendix \ref{sec:equipQuickRef} for image source and archive details. * indicates the data were obtained from our local repository. Table entries with multiple comma-separated values contain groups of exposures taken with different filters.
    \label{tab:observations}
\end{table*}

We identified candidate images where \objname{} was expected to be within the field of view (FOV) based on observation times, pointing centers, and FOV sizes and orientations, downloaded associated data, and extracted image cutouts. We organized data by instrument and observation date and summarize observation details in Table \ref{tab:observations}.

\begin{figure*}
    \centering
    \begin{tabular}{ccccc}
    \begin{overpic}[width=0.18\linewidth]{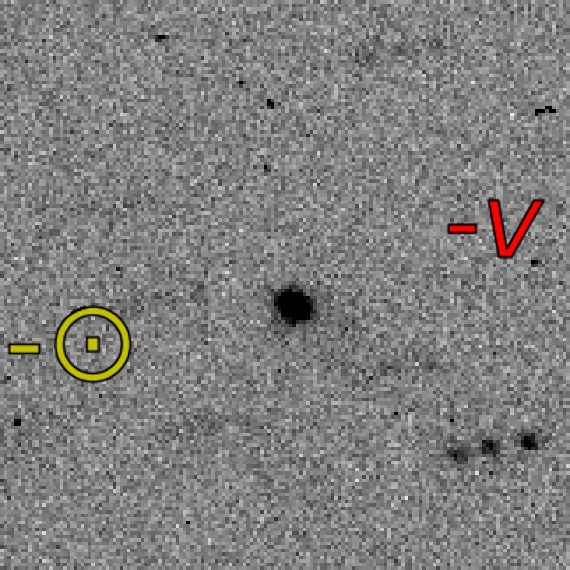}\put (5,7) {\huge\color{green} \textbf{a}}\end{overpic} & 
    \begin{overpic}[width=0.18\linewidth]{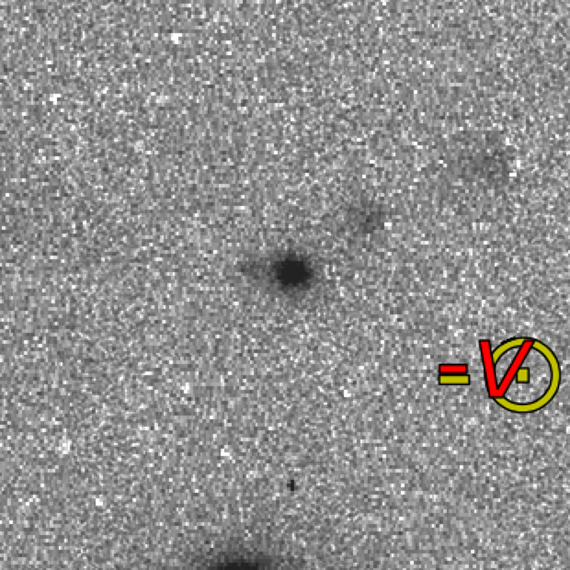}\put (5,7) {\huge\color{green} \textbf{b}}\end{overpic} &
    \begin{overpic}[width=0.18\linewidth]{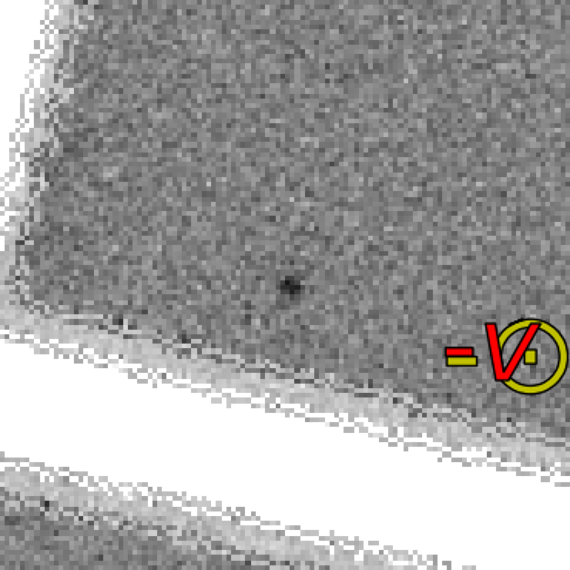}\put (5,7) {\huge\color{green} \textbf{c}}\end{overpic} &
    \begin{overpic}[width=0.18\linewidth]{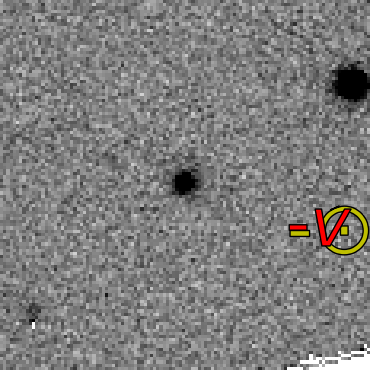}\put (5,7) {\huge\color{green} \textbf{d}}\end{overpic} &
    \begin{overpic}[width=0.18\linewidth]{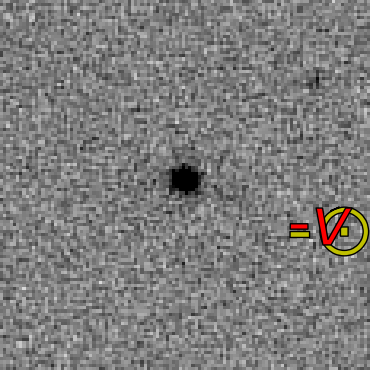}\put (5,7) {\huge\color{green} \textbf{e}}\end{overpic} \\
    \begin{overpic}[width=0.18\linewidth]{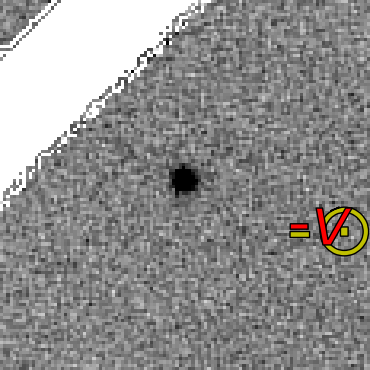}\put (5,7) {\huge\color{green} \textbf{f}}\end{overpic} &
    \begin{overpic}[width=0.18\linewidth]{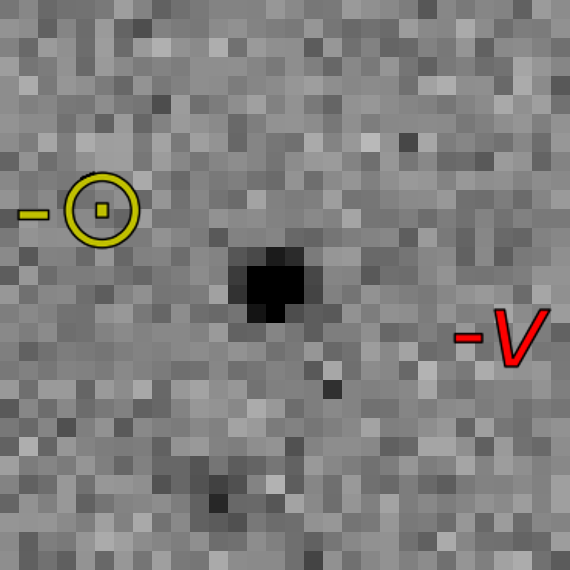} \put (5,7) {\huge\color{green} \textbf{g}}\end{overpic} &
    \begin{overpic}[width=0.18\linewidth]{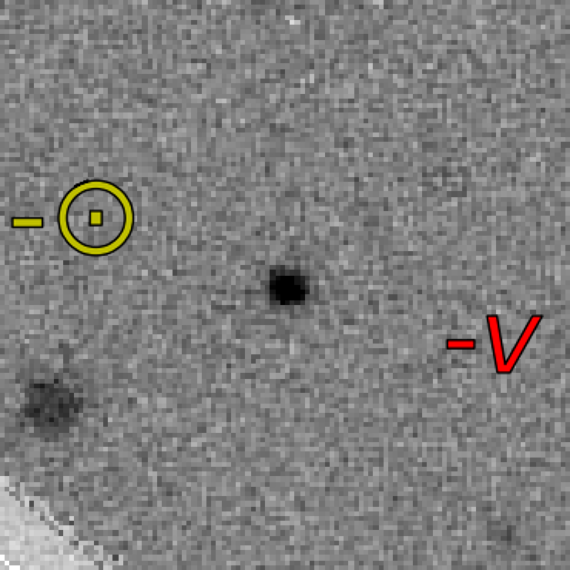}\put (5,7) {\huge\color{green} \textbf{h}}\end{overpic} &
    \begin{overpic}[width=0.18\linewidth]{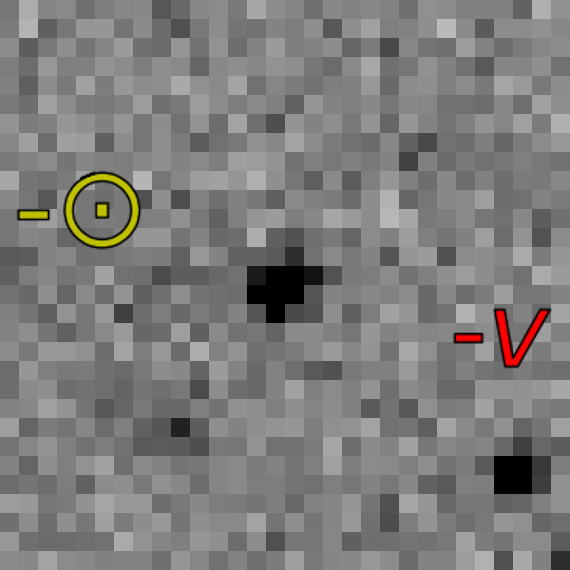} \put (5,7) {\huge\color{green} \textbf{i}}\end{overpic} &
    \begin{overpic}[width=0.18\linewidth]{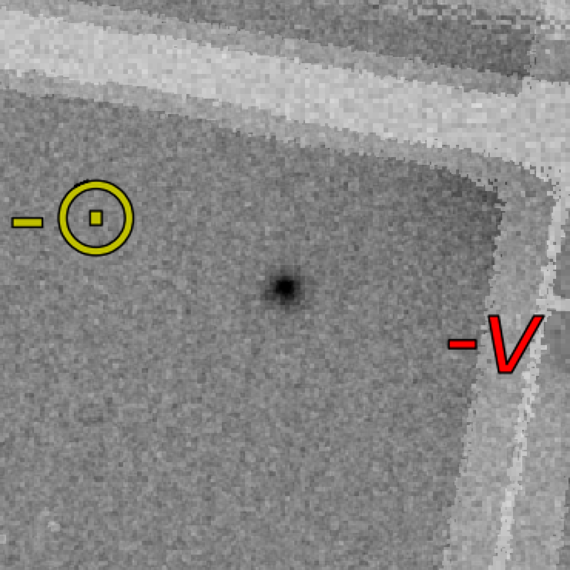} \put (5,7) {\huge\color{green} \textbf{j}}\end{overpic}\\
    \begin{overpic}[width=0.18\linewidth]{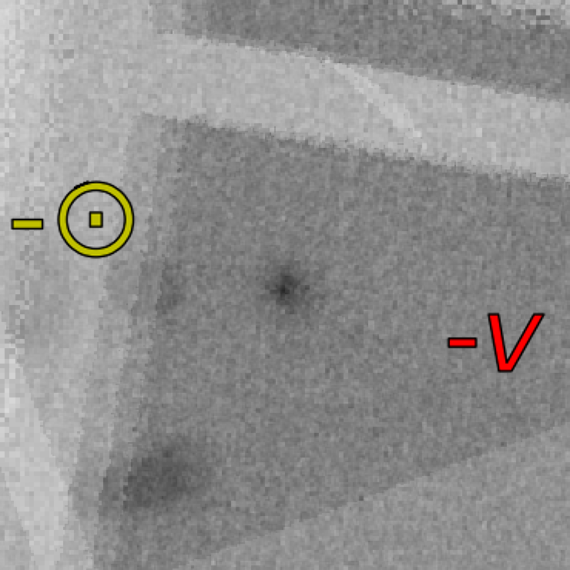} \put (5,7) {\huge\color{green} \textbf{k}}\end{overpic} &
    \begin{overpic}[width=0.18\linewidth]{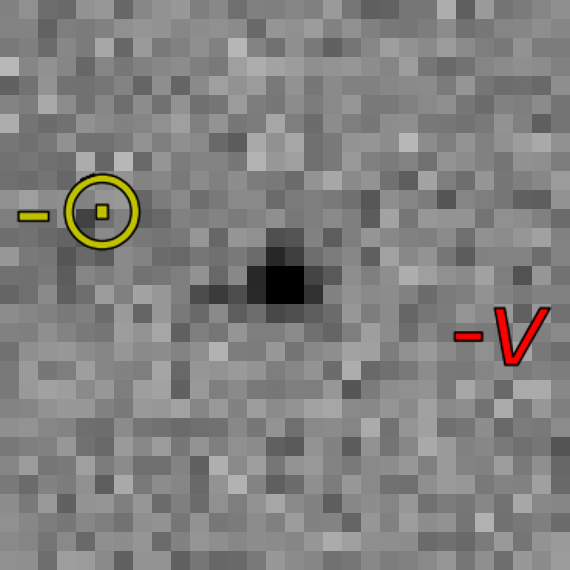} \put (5,7) {\huge\color{green} \textbf{l}}\end{overpic} &
    \begin{overpic}[width=0.18\linewidth]{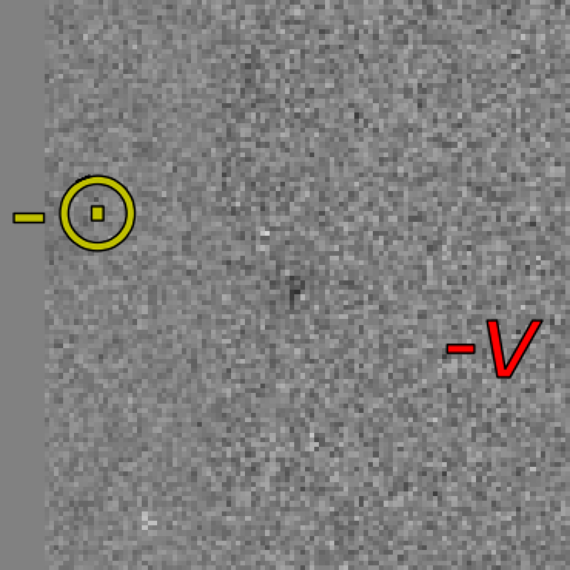} \put (5,7) {\huge\color{green} \textbf{m}}\end{overpic} &
    \begin{overpic}[width=0.18\linewidth]{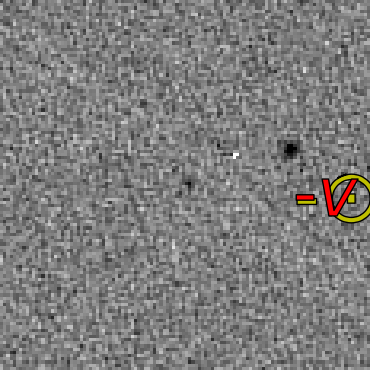} \put (5,7) {\huge\color{green} \textbf{n}}\end{overpic} &
    \begin{overpic}[width=0.18\linewidth]{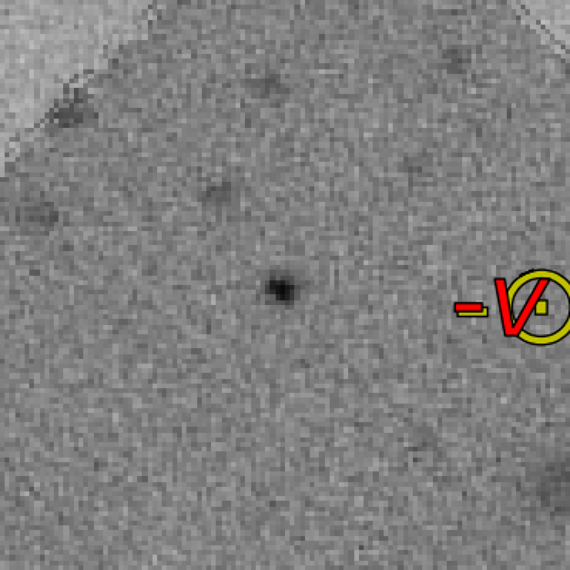}\put (5,7) {\huge\color{green} \textbf{o}}\end{overpic} \\
    \begin{overpic}[width=0.18\linewidth]{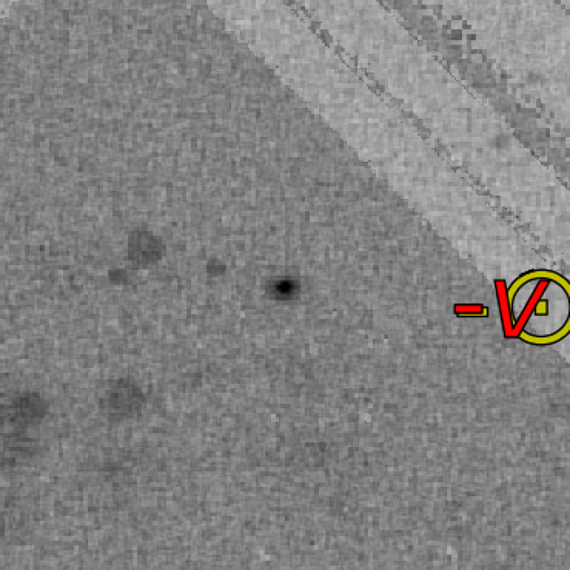}\put (5,7) {\huge\color{green} \textbf{p}}\end{overpic} &
    \begin{overpic}[width=0.18\linewidth]{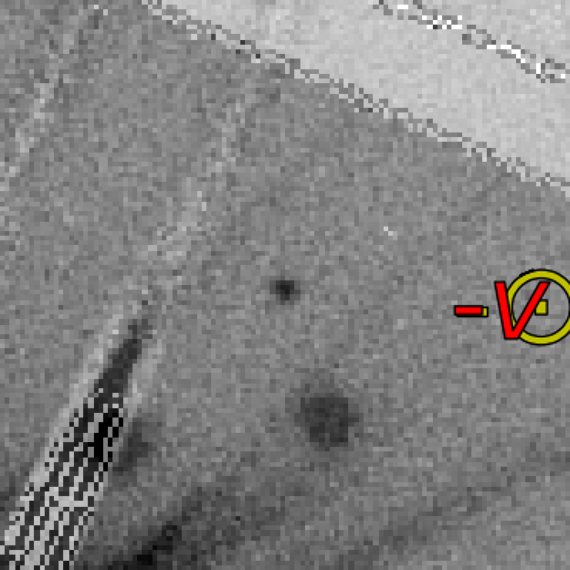}\put (5,7) {\huge\color{green} \textbf{q}}\end{overpic} &
    \begin{overpic}[width=0.18\linewidth]{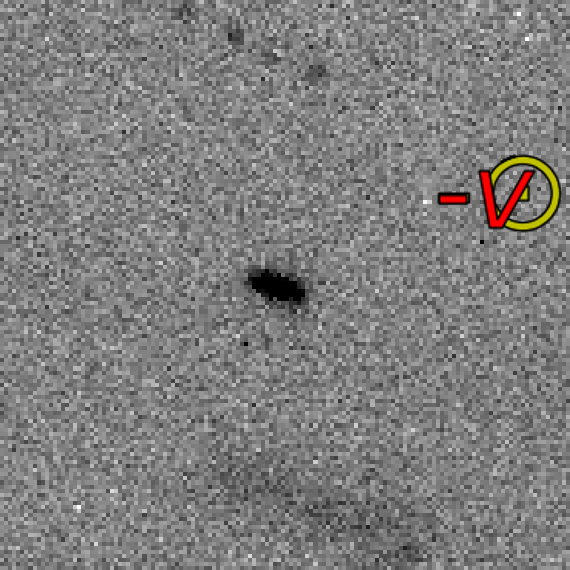}\put (5,7) {\huge\color{green} \textbf{r}}\end{overpic} & 
    \begin{overpic}[width=0.18\linewidth]{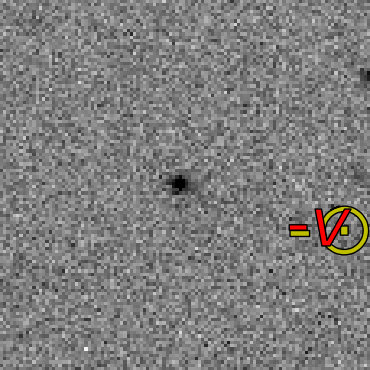}\put (5,7) {\huge\color{green} \textbf{s}}\end{overpic} &
    \begin{overpic}[width=0.18\linewidth]{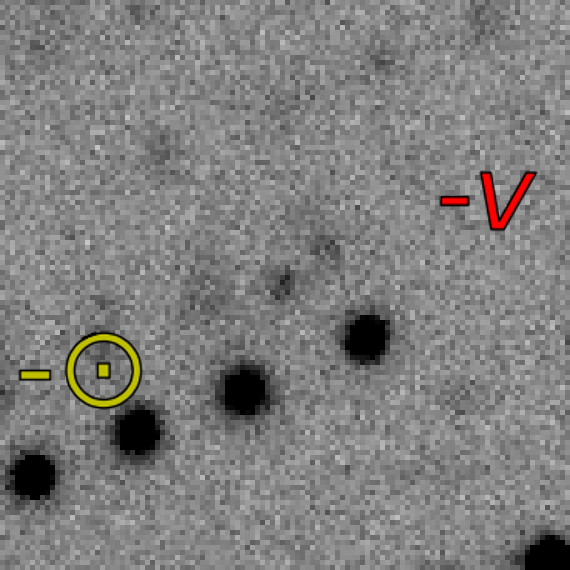}\put (5,7) {\huge\color{green} \textbf{t}}\end{overpic} \\
    \begin{overpic}[width=0.18\linewidth]{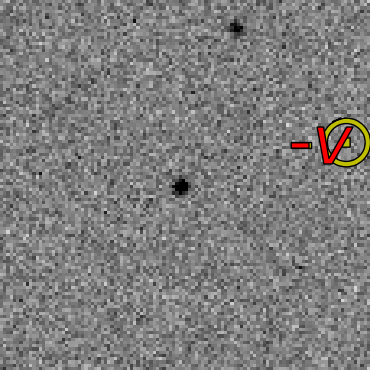}\put (5,7) {\huge\color{green} \textbf{u}}\end{overpic} & 
    \begin{overpic}[width=0.18\linewidth]{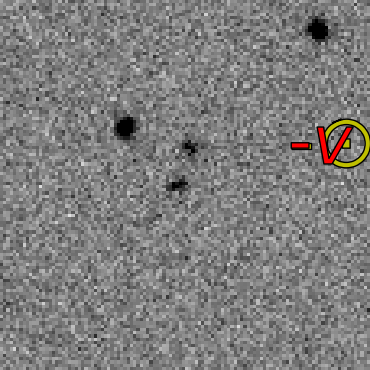}\put (5,7) {\huge\color{green} \textbf{v}}\end{overpic} & 
    \begin{overpic}[width=0.18\linewidth]{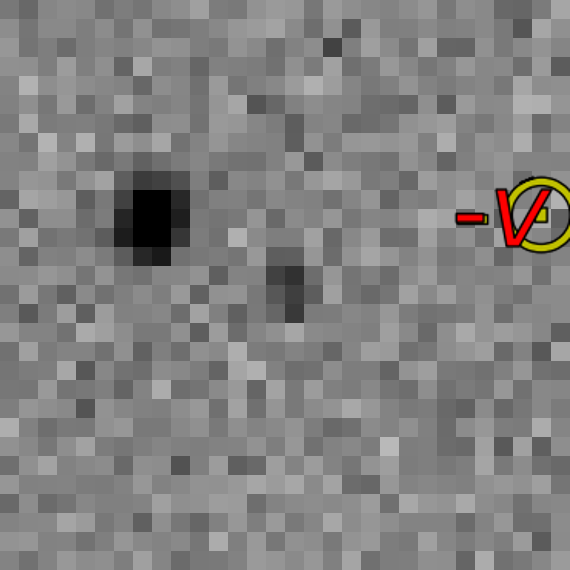}\put (5,7) {\huge\color{green} \textbf{w}}\end{overpic} &
    \begin{overpic}[width=0.18\linewidth]{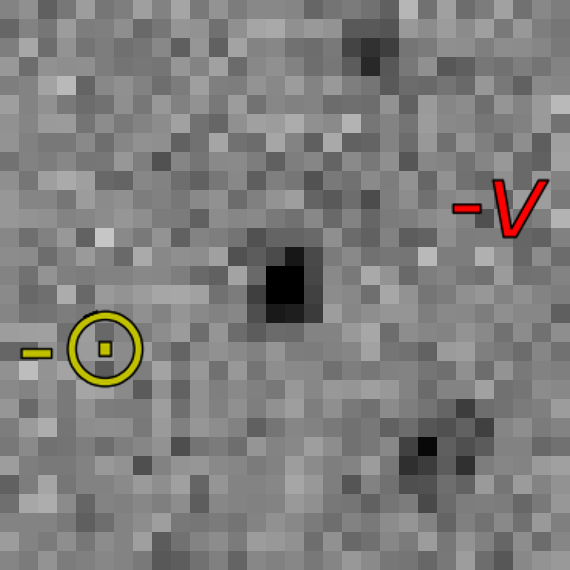}\put (5,7) {\huge\color{green} \textbf{x}}\end{overpic} \\ 
    \end{tabular}
    \caption{Archival images of \objname{} with the best activity detection potential (i.e., sufficient depth and observing conditions) for \objname{}. For all images, north is up, east is left, the FOV is 30\arcsec$\times$30\arcsec, the antisolar (yellow -$\odot$) and antimotion (red -$v$) directions are shown with the origin at image center. See Appendix \ref{sec:equipQuickRef} for instrument and archive details. Panel (s) is from the only thumbnail in which we could identify activity unambiguously (Figure \ref{fig:wedgephot}). (a) 2004 July 8 MegaPrime 3$\times$180~s $i$ band. (b) 2005 June 8 SuprimeCam 3$\times$60~s \textit{W-J-VR} band. (c) 2010 June 14 PS1 2$\times$60~s $z$ band. (d) 2010 August 2 PS1 1$\times$45~s $i$ band. (e) 2010 August 5 PS1 1$\times$40~s $r$ band. (f) 2010 August 6 PS1 1$\times$43~s $g$ band. (g) 2010 August 28 PTF 2$\times$60~s $r$ band. (h) 2010 August 31 PS1 $\times$45~s $i$ band. (i) 2010 September 1 PTF 2$\times$60~s $r$ band. (j) 2010 September 6 PS1 2$\times$43~s $g$ band. (k) 2010 September 12 PS1 2$\times$40~s $r$ band + 2$\times$43~s $g$ band. (l) 2010 September 15 PTF 2$\times$60~s $r$ band. (m) 2010 October 30 PS1 2$\times$30~s $z$ band. (n) 2011 July 14 PS1 1$\times$40~s $r$ band. (o) 2011 November 24 PS1 2$\times$40~s $r$ band + 2$\times$43~s $g$ band. (p) 2011 November 30 PS1 2$\times$45~s $i$ band. (q) 2011 December 1 PS1 2$\times$43~s $g$ band. (r) 2014 January 31 OmegaCAM 5$\times$360~s $r$ band. (s) 2016 July 22 1$\times$89~s \textit{z} band. (t) 2019 July 3 DECam 9$\times$40~s \textit{VR} band. (u) 2020 February 4 DECam 1$\times$38~s \textit{r} band. (v) 2020 February 10 DECam 1$\times$199~s $z$ band. (w) 2020 April 25 ZTF 1$\times$30~s \textit{g} band + 1$\times$30~s \textit{r} band. (x) 2020 May 18, 27 + 2020 June 11, 14, 17, 20, 23, \& 26 ZTF 9$\times$30~s \textit{r} band + 12$\times$30~s \textit{g} band.}
    \label{fig:gallery}
\end{figure*}

We extracted \totalthumbs{} $480\times480$ pixel thumbnail images (such as the image shown in Figure \ref{fig:wedgephot}) in which we could confidently identify \objname{} (Table \ref{tab:observations}). We coadded images from the same instrument when observations were close enough in time for computed tail orientation to be in close agreement such that coaddition could enhance activity, if present. The thumbnails with the ``best activity detection potential'' -- meaning the images were judged to have observing conditions (e.g., seeing) and depth (i.e., magnitude limit) amenable to activity detection -- are shown in Figure \ref{fig:gallery}. 
To allow for uniform spatial comparisons and to magnify the region of interest around \objname{}, all thumbnail images in Figure \ref{fig:gallery} 
are displayed with 30\arcsec$\times$30\arcsec fields of view.

\subsection{Image Assessment}
\label{subsec:imageassessment}

We vetted each thumbnail by visually confirming \objname{} was visible. In cases where the object could not be readily identified, we employed our pipeline to produce comparison thumbnails derived from DECam data that showed the same region of sky, instrument, broadband filter, and exposure time, but from epochs when the object was not in the FOV. We made use of Gaia DR2 \citep{gaiacollaborationGaiaDataRelease2018} and Sloan Digital Sky Survey Release 9 (SDSS~DR-9) catalogs \citep{ahnNinthDataRelease2012} to visually validate World Coordinate System of images within the SAOImageDS9 Vizier \citep{ochsenbeinVizieRDatabaseAstronomical2000} catalog query system.

We next identified vetted thumbnails that were suitable for coaddition by clustering images based on instrument and date. For compatible image sets that included multiple broadband filters, we carried out coaddition among matching filters as well as combining all images. Finally, we visually examined the results and flagged images with potential activity.

We found a single image with clear evidence of activity (Figure \ref{fig:wedgephot}) in an 89~s $z$-band exposure captured 2016 July 22 by Dustin Lang and Alistair Walker as part of the DECam Legacy Survey \citep[DECaLS;][]{deyOverviewDESILegacy2019}. This discovery makes \objname{} the ninth recurrently active main-belt asteroid to be identified to date. The other objects, 238P/Read, 259P/Garradd, 288P, 311P/PANSTARRS, 313P/Gibbs, 324P/La Sagra, (6478)~Gault, and (7968)~Elst-Pizarro, have all demonstrated recurrent activity near perihelion, with the exception of (6478)~Gault \citep{chandlerSixYearsSustained2019}.

We measured the tail length to be about $2\farcm14$ ($2.4\times10^5$~km) in this image but a longer tail may well have been revealed with a longer exposure (see \citet{hsiehPhysicalCharacterizationMainBelt2021} for 2021 apparition tail measurement). Applying our wedge photometry technique (Section \ref{subsec:wedgephotometry}), we produced a diagnostic plot (Figure \ref{fig:wedgephot}) and measured a position angle on the sky of $251\fdg3\pm1\fdg4$ for the tail, in close agreement with the Horizons computed $251\fdg6$ antisolar and $251\fdg7$ antimotion vectors.

\subsection{Wedge Photometry Tail Tool}
\label{subsec:wedgephotometry}

We crafted a new algorithm to (a) identify potentially active objects by detecting likely tails, and (b) quantify alignment between an observed tail and predicted antisolar and antimotion vectors, which are commonly associated with tail direction. Currently, the tool is designed to analyze single tails $<15\degr$ in angular extent, though we plan to address multiple tails and comae in the future. The technique, which we refer to as \textit{wedge photometry}, sums all pixel values in a variable-width wedge bound between an inner and outer radius and identifies wedges containing excess flux relative to other wedges, if present. A similar approach was used in \cite{2011Icar..215..534S} but we have made improvements in angular resolution and algorithmic efficiency. Excess flux within a particular wedge may indicate the presence of a tail, and testing tail alignment with antisolar and antimotion vectors provides additional weight that a detected tail is real. Here we focused on quantifying tail orientation and position angle agreement.

To optimize the process, we convert Cartesian pixel coordinates ($x$,$y$) to polar coordinates ($r$,$\theta$) where the central thumbnail pixel is defined as (0,0). The resulting three-dimensional array has columns $r$, $\theta$, and $c$ (counts).

For a series of wedge sizes $\theta$ ($1\degr$--$10\degr$ in $1\degr$ increments) we summed pixel values in annular segments spanning an angle $\pm\theta/2$ along a radial component $r$ from an inner bound, $r_0=5$ pixels, to a maximum of $r_\mathrm{max}=50$ pixels (13\arcsec for our DECam data), as given by

\begin{equation}
    c(\theta, \Delta\theta) = \sum_{\theta=-\Delta\theta/2}^{\theta=+\Delta\theta/2} \sum_{r=r_0}^{r=r_\mathrm{max}} c(r, \theta).
\end{equation}

\noindent We further optimize the procedure by selecting for the target a starting radius $r_0$ outside the FWHM, and choosing a maximum radius $_\mathrm{max}$ that allows for a wedge length long enough to ensure that all bins have sufficient counts to avoid necessitating resampling of any individual pixels. Thus, pixels are assigned to wedges based solely on their precise pixel center coordinate, and any fractional flux from a pixel that spans a wedge boundary is automatically assigned to the wedge containing the pixel center coordinate. We then compute the mean and standard deviation of the resulting counts for each $\theta$ to compare with the predicted antisolar and antimotion vectors.

We produce a polar plot (Figure \ref{fig:wedgephot}) to aid assessing relative radial flux distribution. Most position angles have a $1\sigma$ of $\sim$200 counts. The tail is clearly identified by our algorithm at $> 7 \sigma$ for several $\Delta \theta$ wedge sizes.


\section{Main-belt Comet Classification}
\label{sec:mainBeltComet}

Once we had identified a previous activity epoch we set out to determine if \objname{} could be an MBC.

\subsection{Prerequisites}
\label{subsec:mbc}
For \objname{} to qualify as an MBC it must (1) be an active asteroid, (2) orbit within the Main Asteroid Belt, and (3) exhibit sublimation driven activity.

(1) To qualify as an active asteroid, a body must typically meet three criteria (see \citealt{jewittActiveAsteroids2012} for discussion): (i) A coma or tail must have been observed visually (as is the case in this work) or, potentially, through alternate means such as spectroscopy \citep[e.g.,][]{kuppersLocalizedSourcesWater2014,busarevNewCandidatesActive2018} or detecting magnetic field enhancements \citep[e.g.,][]{russellInterplanetaryMagneticField1984}. (ii) The semi-major axis $a$ must not be exterior to that of Jupiter ($a_\mathrm{J}\approx$ 5.2~au) as is the case for comets and active Centaurs \citep{jewittActiveCentaurs2009}; \objname{} has $a=$3.1~au. And (iii) the Tisserand parameter with respect to Jupiter, $T_\mathrm{J}$, must be greater than 3; this is because objects with $T_\mathrm{J}<3$ are canonically considered cometary and $T_\mathrm{J}>3$ asteroids \citep{vaghiOriginJupiterFamily1973,vaghiOrbitalEvolutionComets1973}.

$T_\mathrm{J}$ describes how an orbit is related to Jupiter by 

\begin{equation}
    T_\mathrm{J} = \frac{a_\mathrm{J}}{a} + 2 \sqrt{\frac{a\left(1-e^2\right)}{a_\mathrm{J}}}\cos\left(i\right).
\end{equation}

\noindent where $e$ is the eccentricity and $i$ is the orbital inclination. $T_\mathrm{J}$ for \objname{} is 3.192 and thus it qualifies as asteroidal. \objname{} properties are provided in Appendix \ref{sec:ObjectData}, and are established with this criterion.

(2) \objname{} orbits between 2.4~au and 3.76~au and thus does not cross the orbits of either Mars or Jupiter. With a semi-major axis of 3.1~au, \objname{} is an outer main-belt asteroid orbiting between the Kirkwood gaps corresponding to the 7:8 and 2:1 mean motion resonances with Jupiter.

(3) Recurrent activity near perihelion is diagnostic of volatile sublimation as the most likely mechanism responsible for the observed activity  \citep[e.g.,][]{hsiehOpticalDynamicalCharacterization2012}. However, other underlying causes of recurrent activity are known, so this point warrants further investigation.

\subsection{Activity Mechanism}
\label{subsec:mechanism}

We demonstrated in Section \ref{sec:secondActivityEpoch} that \objname{} has been active during at least two epochs. This helps rule out activity mechanisms  such as \textit{impact events} (e.g., (596)~Scheila; \citealt{bodewitsCollisionalExcavationAsteroid2011,ishiguroObservationalEvidenceImpact2011,moreno596ScheilaOutburst2011}) that are only expected to produce one-time outbursts but which can expel dust and produce comet-like activity. Aside from \textit{temperature-correlated volatile sublimation} (which we examine further in Section~\ref{subsec:tempestimation}) other mechanisms for producing recurrent activity have been proposed.

\textit{Rotational destabilization} causes dust to be flung from a body in a potentially multiepisodic manner, as may be the case for (6478)~Gault \citep{chandlerSixYearsSustained2019,kleynaSporadicActivity64782019}. Taxonomic classification can help diagnose rotational destabilization, as with $S$-type (6478)~Gault, because activity from desiccated asteroid classes is unlikely to be sublimation driven. As discussed in Section \ref{subsec:tempestimation}, the taxonomic class of \objname{} is not yet known but it is likely a C-type asteroid. An accurate rotation period for \objname{} is currently unavailable, and as such, we can neither confirm nor rule out destabilization as a contributing factor to the observed activity at this time.

\textit{Rubbing binaries} is a hypothetical scenario whereby two merging asteroids repeatedly collide and eject material. Proposed as a possible mechanism for the activity of 311P/PANSTARRS \citep{hainautContinuedActivity20132014}, the rubbing binary scenario has yet to be confirmed for that object \citep{jewittNucleusActiveAsteroid2018} or any other. As of this writing, there is no evidence that \objname{} is a binary asteroid, and activity spans two epochs separated by 5 yr, so we would expect merging processes to have either finished or that the binary orbit would have stabilized (see \citealt{jewittNucleusActiveAsteroid2018} for additional discussion concerning dissipation timescales). Therefore we find it unlikely that rubbing causes the activity associated with \objname{}.

Geminid meteor stream parent (3200)~Phaethon undergoes extreme temperature changes ($\sim$600~K) and peaks at 800~K to 1100~K, well above the 573~K serpentine-phyllosilicate decomposition threshold \citep{ohtsukaSolarRadiationHeatingEffects2009}. These temperatures likely induce \textit{thermal fracture} \citep{licandroNatureCometasteroidTransition2007,kasugaObservations1999YC2008} leading to mass shedding \citep{liRecurrentPerihelionActivity2013,huiResurrection3200Phaethon2017}.

Two mechanisms, thermal fracture and temperature-correlated volatile sublimation warrant further inquiry into the thermophysical properties of \objname{}.

\subsection{Temperature Estimation}
\label{subsec:tempestimation}

Estimating temperatures experienced by \objname{} aids us in understanding direct thermal effects (e.g., thermal fracture) as well as assessing long-term volatile survival, especially water. For these reasons, we computed temperatures for an airless body over the course of an orbit similar to that of \objname{}.

Following \citet{hsiehMainbeltCometsPanSTARRS12015}, the energy balance equation for a gray body on which water ice sublimation is occurring is 
\begin{equation}
{F_{\odot}\over r_h^2}(1-A) = \chi\left[{\varepsilon\sigma T_{eq}^4 + L f_D\dot m_{w}(T_{eq})}\right]
\label{equation:sublim1}
\end{equation}
where $r_h$ is the object's heliocentric distance, $T_\mathrm{eq}$ is the equilibrium surface temperature, $F_{\odot}=1360$~W~m$^{-2}$ is the solar constant, $r_h$ is in au, $A=0.05$ is the assumed Bond albedo of the body, $\chi$ accounts for the distribution of solar heating over the object's surface, $\sigma$ is the Stefan--Boltzmann constant, and $\varepsilon=0.9$ is the assumed effective infrared emissivity, and $L=2.83$~MJ~kg$^{-1}$ is the latent heat of sublimation of water ice (which we approximate here as being independent of temperature), $f_D$ represents the reduction in sublimation efficiency caused by mantling, where $f_D=1$ in the absence of a mantle, and $\dot m_w$ is the water mass-loss rate due to sublimation of surface ice.

In this equation, $\chi=1$ corresponds to a flat slab facing the Sun, known as the subsolar approximation, and produces the maximum expected temperature for an object, 
while $\chi=4$ applies to objects with fast rotation or low thermal inertia, known as the isothermal approximation, and produces the minimum expected temperature for an object.

Next, the sublimation rate of ice into a vacuum can be computed using
\begin{equation}
\dot m_{w} = P_v(T) \sqrt{\mu\over2\pi k T}
\label{equation:sublim2}
\end{equation}
where $\mu=2.991\cdot 10^{-26}$~kg is the mass of one water molecule, and $k$ is the Boltzmann constant, and the equivalent ice recession rate, $\dot \ell_{i}$, corresponding to $\dot m_{w}$ is given by
$\dot \ell_{i} = \dot m_{w}/ \rho$,
where $\rho$ is the bulk density of the object.

Finally, the Clausius--Clapeyron relation,
\begin{equation}
P_v(T) = 611 \times \exp\left[{{\Delta H_\mathrm{subl}\over R_g}\left({{1\over 273.16} - {1\over T}}\right)}\right]
\label{equation:sublim3}
\end{equation}
gives the vapor pressure of water, $P_v(T)$, in Pa, where $\Delta H_\mathrm{subl}=51.06$~MJ~kmol$^{-1}$ is the heat of sublimation for ice to vapor and $R_g=8314$~J~kmol$^{-1}$~K$^{-1}$ is the ideal gas constant.
Solving these three equations iteratively, one can calculate the equilibrium temperature of a gray body at a given heliocentric distance on which water ice sublimation is occurring.

\begin{figure*}[ht]
	\centering
	\begin{tabular}{c}
	        \hspace{-4mm}\includegraphics[width=0.85\linewidth]{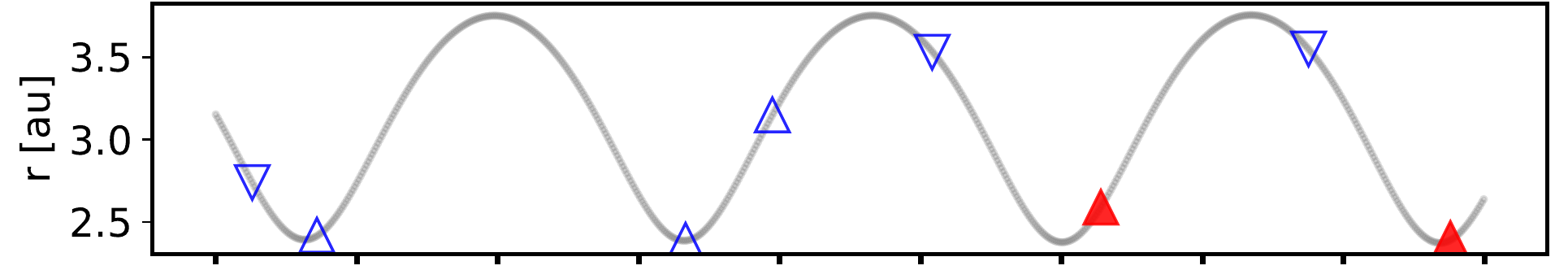}\\
	     	\hspace{9mm}\includegraphics[width=0.92\linewidth]{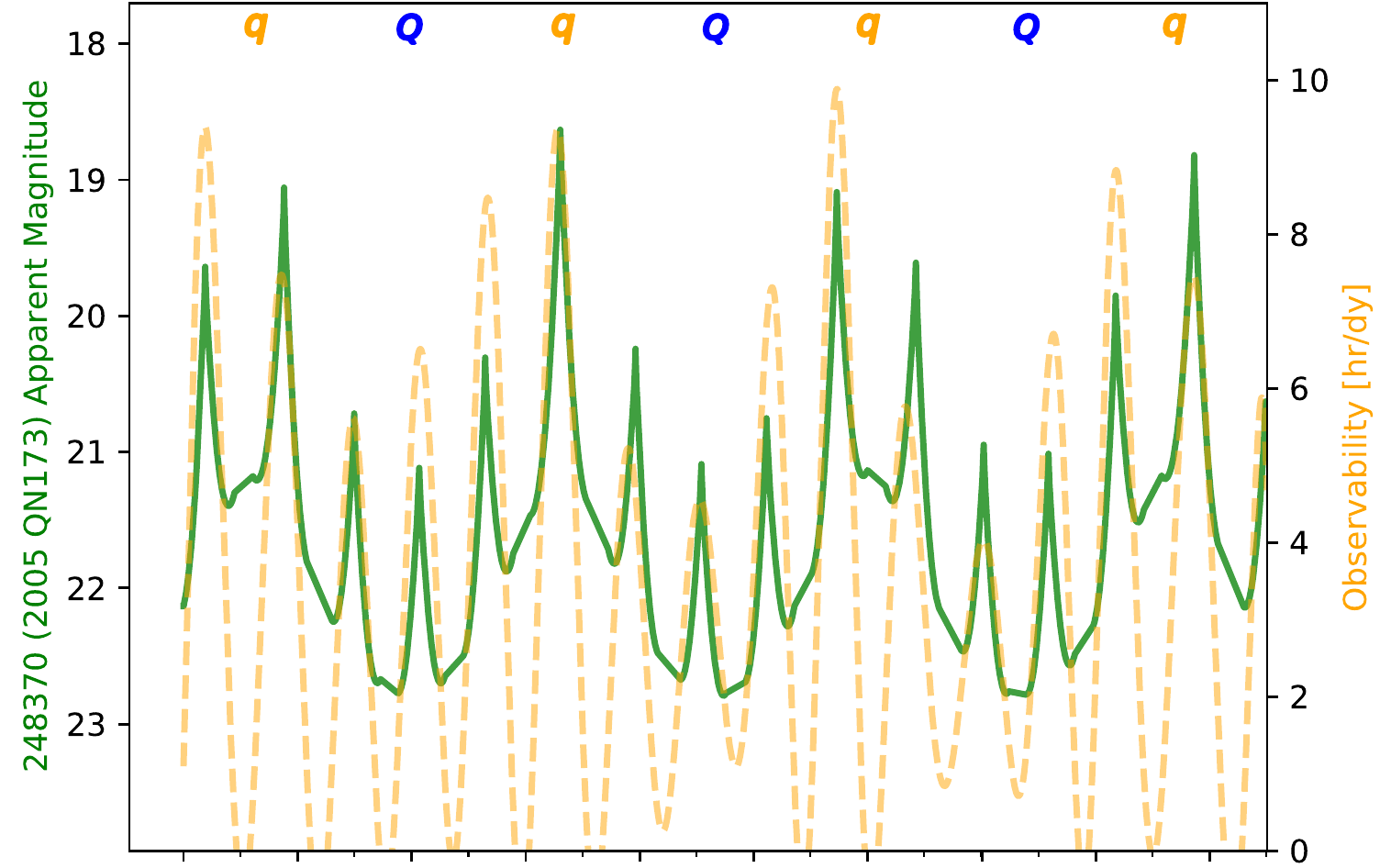}\\
	       \hspace{-4mm}\includegraphics[width=0.845\linewidth]{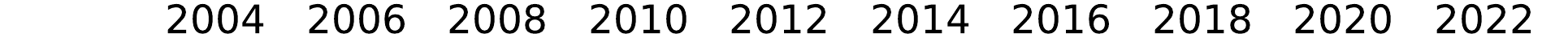}\\
	       \hspace{-1mm}\includegraphics[width=0.845\linewidth]{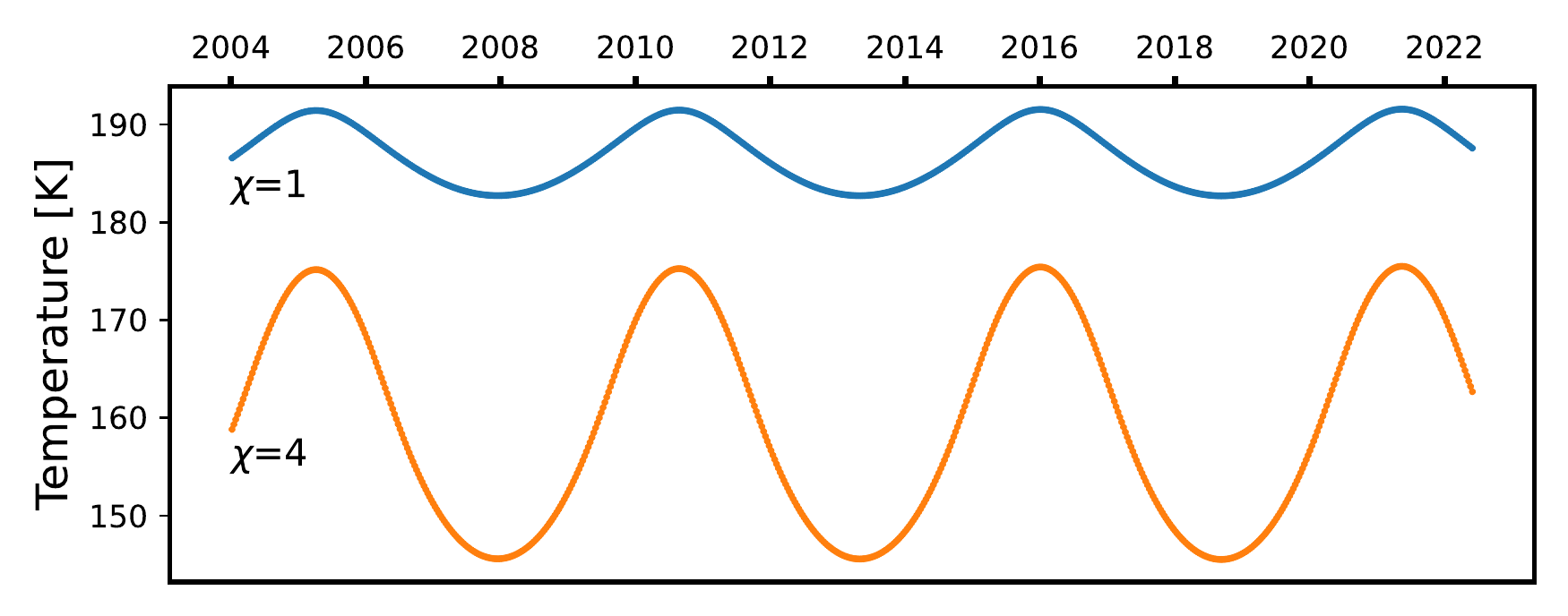}
	\end{tabular}
	\caption{\objname{} heliocentric distance (top plot), observability timeline (middle plot) and temperature (bottom plot), beginning the year of our first archival data (2004) through 2022.
	Top: triangles represent positive (filled red) and negative (unfilled blue) activity detections. Markers indicate when the object was inbound (downward pointing triangles) or outbound (upward pointing triangles). Table \ref{tab:observations} lists observation details. 
	Middle: apparent $V$-band magnitude (solid green line) and our ``observability'' metric (yellow dashed line) that represents hours during a given UT observing date the object is above $> 15\degr$ elevation while the Sun is below the horizon. Peaks in apparent magnitude coinciding with observability occur during opposition events, and observability troughs indicate solar conjunctions when \objname{} was only above the horizon during daylight. Perihelion (orange $q$) and aphelion (blue $Q$) events are indicated. 
	Bottom: temperature $T$~(K) by date for different $\chi$ values, where $\chi=1$ (top line) represents a ``flat slab'' and $\chi=4$ (bottom line) an isothermal body.
    }
	\label{fig:ActivityTimeline}
\end{figure*}

In Figure~\ref{fig:ActivityTimeline}, we plot the object's expected equilibrium temperature over several orbit cycles, as computed by solving the system of equations above. We plot temperatures computed using both $\chi=1$ and $\chi=4$ to show the full range of possible temperatures.

%
%

Figure~\ref{fig:ActivityTimeline} (top) shows two parameters key to observing \objname{} between 2015 and 2023: apparent $V$-band magnitude and ``observability,'' which we define as the number of hours an object remains above $15\degr$ elevation during nighttime for a given UT observing date. This plot informs us of potential observational biases or geometric effects that may bias activity detection, such as preferential activity discovery during opposition events, as was the case with (6478)~Gault \citep{chandlerSixYearsSustained2019}.

Figure~\ref{fig:ActivityTimeline} (bottom) illustrates \objname{}'s heliocentric distance and temperature (as computed using Equations \ref{equation:sublim1}-\ref{equation:sublim3}) over time, plus dates of observed activity and images where no visible activity was conclusively identified. Throughout its entire orbit the surface of \objname{} is consistently warmer than 145~K, the temperature above which water ice is not expected to survive over timescales on the order of the age of the solar system \citep{schorghoferLifetimeIceMain2008,snodgrassMainBeltComets2017}. 

However, it is possible for water ice to remain preserved over long (Gyr) timescales on small asteroids at depths as shallow as a few centimeters to 30~cm below the surface \citep{schorghoferLifetimeIceMain2008,prialnikCanIceSurvive2009}, where present-day activity may be triggered by meter-scale impactors that excavate subsurface ice. We note that water ice has been detected on the surface of main-belt asteroid (24) Themis \citep{campinsWaterIceOrganics2010,rivkinDetectionIceOrganics2010}, but the mechanism by which that water ice is able to persist on its surface -- likely requiring continual replenishment from subsurface volatile reservoirs -- is not well understood, and furthermore may not have the same effectiveness on kilometer-scale objects like \objname{} as it does on the 200~km diameter (24) Themis.

The surface temperature of \objname{} varies at most between 145~K and 190~K over its orbit (Figure \ref{fig:ActivityTimeline}), far less than the 600~K temperature swings peaking at 800--1000~K described in Section \ref{subsec:mechanism}. We consider it is unlikely that thermal fracture is the primary cause of \objname{}'s activity.

\subsection{Nondetection of Activity}
\label{subsec:nullresults}

The two known epochs of activity for \objname{} both occurred when the object was interior to a heliocentric distance of 2.7~au. However, \objname{} was observed in 2005 and 2010 when the object was interior to 2.7~au but no activity was detected. We believe the circumstances of these epochs preclude a definitive assessment of activity.

The 2010 Pan-STARRS1 data suffer from image artifacts that are significant enough to obscure activity that may have been present. The 2005 SuprimeCam observations should have been well suited to detecting activity as the 8~m Subaru telescope has a large aperture, exposure times (60~s) were sufficiently long, the \textit{W-J-VR} filter covered a broad wavelength range, and the object was well placed in the sky in terms of airmass/elevation during the observations. However, extinction varied significantly over the observing sequence as the summit log for that night\footnote{\url{https://smoka.nao.ac.jp/calendar/slog/2005/slog_20050608.txt}} indicated that conditions were windy with cirrus clouds and the differential image motion monitor measured significant seeing variation (roughly $0\farcs8$ to $>3\arcsec{}$)\footnote{\url{https://smoka.nao.ac.jp/calendar/subaruseeing/20050608.gif}}, potentially contributing to a considerable reduction in our ability to detect activity. All sources in the field could be matched to SDSS~DR-9 stars but the faintest stars we were able match to the SDSS~DR-9 catalog were $r\approx21.3$, very similar to the JPL Horizons computed $V$=21.0 for \objname{}. The Subaru Exposure Time Calculator estimates an equivalent $r$-band exposure would deliver a signal-to-noise ratio of 66 for a source of equivalent magnitude, but we estimate the images are at best $\sim$0.5 mag deeper than necessary to detect the object and thus we find it unlikely that activity would be detectable unless the object was undergoing a significant outburst at the time.

Although we cannot definitively rule out the presence of activity in 2005 or 2010 from these observations, another possibility is that a triggering event (e.g., impact, rotational destabilization) that started \objname{}'s current activity occurred between 2005 June and 2016 July. This would explain the object's apparent inactivity and activity on each of those dates, respectively.

\subsection{Main-belt Comet Membership}
\label{subsec:mbcmembership}
Given the above reasoning, we find it most likely that the activity associated with \objname{} is sublimation driven, in which case the object is an MBC. However, we emphasize that in order to rule out rotational destabilization as the root cause of the observed activity, additional work is needed. Moreover, confirmation of a third activity epoch would lend further evidence favoring sublimation as the primary agency of activity.

\section{Summary and Future Work}
\label{sec:summary}
We harvested \totalthumbs{} images of \objname{} (also designated 433P) spanning \totalobservations{} observing dates. We found clear evidence of a previous activity epoch on 2016 July 22. We provide a catalog of archival observations along with an image gallery. Making use of wedge photometry -- a novel tail detection and characterization tool we introduce in this Letter -- we measure tail orientation to be in close agreement with the antisolar and antimotion vectors computed by Horizons. We showed that \objname{} is a likely member of the MBCs, a group of active asteroids orbiting within the Main Asteroid Belt that are active due to volatile sublimation.

The current observing window for this object ends around 2021 December, and when it returns in late 2022 it will be over 3~au from the Sun and less likely to show activity. We did not find any images showing \objname{} active at beyond 3~au, so we call on observers to make use of the present activity apparition while it is still possible to do so. Continued monitoring to study the evolution of the tail's brightness, including surface brightness measurements, can lead to better characterization of ejected dust grain sizes and total mass loss during this apparition. Once activity subsides, time-series observations to measure a rotation period will be especially useful for diagnosing rotational breakup. Preliminary color measurements suggest \objname{} is a C-type asteroid \citep{hsiehPhysicalCharacterizationMainBelt2021}, but a robust taxonomic classification would help further solidify our assessment of the underlying activity mechanism.


\section{Acknowledgements}
\label{sec:acknowledgements}

The authors thank the anonymous referee whose comments greatly improved the quality of this Letter.

We thank Dr.\ Mark Jesus Mendoza Magbanua (University of California San Francisco) for his frequent and timely feedback on the project. 
The authors express their gratitude to 
Prof. Mike Gowanlock (NAU), 
Dr. Annika Gustafsson (NAU, Lowell Observatory, Southwest Research Institute), 
Jay Kueny (Steward Observatory), 
and the Trilling Research Group (NAU), all of whom provided invaluable insights which substantially enhanced this work. The unparalleled support provided by Monsoon cluster administrator Christopher Coffey (NAU) and the High Performance Computing Support team facilitated the scientific process.

This material is based upon work supported by the National Science Foundation Graduate Research Fellowship Program under grant No.\ (2018258765). Any opinions, findings, and conclusions or recommendations expressed in this material are those of the author(s) and do not necessarily reflect the views of the National Science Foundation.  C.O.C., H.H.H. and C.A.T. also acknowledge support from the NASA Solar System Observations program (grant 80NSSC19K0869). 

Computational analyses were run on Northern Arizona University's Monsoon computing cluster, funded by Arizona's Technology and Research Initiative Fund. This work was made possible in part through the State of Arizona Technology and Research Initiative Program. 
World Coordinate System (WCS) corrections facilitated by the \textit{Astrometry.net} software suite \citep{langAstrometryNetBlind2010}.

This research has made use of data and/or services provided by the International Astronomical Union's Minor Planet Center. 
This research has made use of NASA's Astrophysics Data System. 
This research has made use of The Institut de M\'ecanique C\'eleste et de Calcul des \'Eph\'em\'erides (IMCCE) SkyBoT Virtual Observatory tool \citep{berthierSkyBoTNewVO2006}. 
This work made use of the {FTOOLS} software package hosted by the NASA Goddard Flight Center High Energy Astrophysics Science Archive Research Center. 
This research has made use of SAOImageDS9, developed by Smithsonian Astrophysical Observatory \citep{joyeNewFeaturesSAOImage2006}. 
This work made use of the Lowell Observatory Asteroid Orbit Database \textit{astorbDB} \citep{bowellPublicDomainAsteroid1994,moskovitzAstorbDatabaseLowell2021}. 
This work made use of the \textit{astropy} software package \citep{robitailleAstropyCommunityPython2013}.

This project used data obtained with the Dark Energy Camera (DECam), which was constructed by the Dark Energy Survey (DES) collaboration. Funding for the DES Projects has been provided by the U.S. Department of Energy, the U.S. National Science Foundation, the Ministry of Science and Education of Spain, the Science and Technology Facilities Council of the United Kingdom, the Higher Education Funding Council for England, the National Center for Supercomputing Applications at the University of Illinois at Urbana-Champaign, the Kavli Institute of Cosmological Physics at the University of Chicago, Center for Cosmology and Astro-Particle Physics at the Ohio State University, the Mitchell Institute for Fundamental Physics and Astronomy at Texas A\&M University, Financiadora de Estudos e Projetos, Funda\c{c}\~{a}o Carlos Chagas Filho de Amparo, Financiadora de Estudos e Projetos, Funda\c{c}\~ao Carlos Chagas Filho de Amparo \`{a} Pesquisa do Estado do Rio de Janeiro, Conselho Nacional de Desenvolvimento Cient\'{i}fico e Tecnol\'{o}gico and the Minist\'{e}rio da Ci\^{e}ncia, Tecnologia e Inova\c{c}\~{a}o, the Deutsche Forschungsgemeinschaft and the Collaborating Institutions in the Dark Energy Survey. The Collaborating Institutions are Argonne National Laboratory, the University of California at Santa Cruz, the University of Cambridge, Centro de Investigaciones En\'{e}rgeticas, Medioambientales y Tecnol\'{o}gicas–Madrid, the University of Chicago, University College London, the DES-Brazil Consortium, the University of Edinburgh, the Eidgen\"ossische Technische Hochschule (ETH) Z\"urich, Fermi National Accelerator Laboratory, the University of Illinois at Urbana-Champaign, the Institut de Ci\`{e}ncies de l'Espai (IEEC/CSIC), the Institut de Física d'Altes Energies, Lawrence Berkeley National Laboratory, the Ludwig-Maximilians Universit\"{a}t M\"{u}nchen and the associated Excellence Cluster Universe, the University of Michigan, the National Optical Astronomy Observatory, the University of Nottingham, the Ohio State University, the University of Pennsylvania, the University of Portsmouth, SLAC National Accelerator Laboratory, Stanford University, the University of Sussex, and Texas A\&M University.

Based on observations at Cerro Tololo Inter-American Observatory, National Optical Astronomy Observatory (NOAO Prop. ID 2016A-0190, PI: Dey), which is operated by the Association of Universities for Research in Astronomy (AURA) under a cooperative agreement with the National Science Foundation. This research has made use of the NASA/IPAC Infrared Science Archive, which is funded by the National Aeronautics and Space Administration and operated by the California Institute of Technology.

The Legacy Surveys consist of three individual and complementary projects: the Dark Energy Camera Legacy Survey (DECaLS; Proposal ID \#2014B-0404; PIs: David Schlegel and Arjun Dey), the Beijing-Arizona Sky Survey (BASS; NOAO Prop. ID \#2015A-0801; PIs: Zhou Xu and Xiaohui Fan), and the Mayall z-band Legacy Survey (MzLS; Prop. ID \#2016A-0453; PI: Arjun Dey). DECaLS, BASS and MzLS together include data obtained, respectively, at the Blanco telescope, Cerro Tololo Inter-American Observatory, NSF's NOIRLab; the Bok telescope, Steward Observatory, University of Arizona; and the Mayall telescope, Kitt Peak National Observatory, NOIRLab. The Legacy Surveys project is honored to be permitted to conduct astronomical research on Iolkam Du'ag (Kitt Peak), a mountain with particular significance to the Tohono O'odham Nation. BASS is a key project of the Telescope Access Program (TAP), which has been funded by the National Astronomical Observatories of China, the Chinese Academy of Sciences (the Strategic Priority Research Program ``The Emergence of Cosmological Structures'' Grant \# XDB09000000), and the Special Fund for Astronomy from the Ministry of Finance. The BASS is also supported by the External Cooperation Program of Chinese Academy of Sciences (Grant \# 114A11KYSB20160057), and Chinese National Natural Science Foundation (Grant \# 11433005). The Legacy Survey team makes use of data products from the Near-Earth Object Wide-field Infrared Survey Explorer (NEOWISE), which is a project of the Jet Propulsion Laboratory/California Institute of Technology. NEOWISE is funded by the National Aeronautics and Space Administration. The Legacy Surveys imaging of the DESI footprint is supported by the Director, Office of Science, Office of High Energy Physics of the U.S. Department of Energy under Contract No. DE-AC02-05CH1123, by the National Energy Research Scientific Computing Center, a DOE Office of Science User Facility under the same contract; and by the U.S. National Science Foundation, Division of Astronomical Sciences under Contract No. AST-0950945 to NOAO.

Based in part on data collected at Subaru Telescope and obtained from the SMOKA, which is operated by the Astronomy Data Center, National Astronomical Observatory of Japan \citep{2002ASPC..281..298B}.

\vspace{5mm}
\facilities{Astro Data Archive, Blanco (DECam), CFHT (MegaCam), Gaia, IRSA, PO:1.2m (PTF, ZTF), PS1, Sloan, VST (OmegaCAM)}

\software{{\tt astropy} \citep{robitailleAstropyCommunityPython2013},
        {\tt astrometry.net} \citep{langAstrometryNetBlind2010},
        {\tt FTOOLS},
        {\tt JPL Horizons} \citep{Giorgini1996Horizons},
        {\tt SAOImageDS} \citep{joyeNewFeaturesSAOImage2006},
        {\tt SkyBot} \citep{berthierSkyBoTNewVO2006},
        {\tt Vizier} \citep{ochsenbeinVizieRDatabaseAstronomical2000}
          }

\appendix

\section{Equipment and Archives}
\label{sec:equipQuickRef}

\begin{table}
\caption{Equipment and Archives}
\centering
\footnotesize
    \begin{tabular}{llclcccccc}
Instrument  & Telescope            & Pixel Scale    & Location                & NOIR         & ESO & IRSA         & SMOKA & SSOIS        & STScI  \\
            &                       & [\arcsec/pix] &                   &   & & &\\
\hline
\hline
DECam       & 4 m Blanco           & 0.263          & Cerro Tololo, Chile     & S,R &     &              &       & S            &        \\
OmegaCAM    & 2.6 m VLT Survey     & 0.214          & Cerro Paranal, Chile    &              & R   &              &       & S            &        \\
GigaPixel1  & 1.8 m Pan-STARRS1    & 0.258          & Haleakalā, Hawaii       &              &     &              &       & S            & R      \\
MegaPrime   & 3.6 m CFHT           & 0.185          & Mauna Kea, Hawaii       &              &     &              &       & S,R &        \\
PTF/CFHT 12K& 48" Samuel Oschin    & 1.010          & Mt. Palomar, California &              &     & S,R &       &              &        \\
SuprimeCam  & 8.2 m Subaru         & 0.200          & Mauna Kea, Hawaii       &              &     &              & R     & S            &        \\
ZTF Camera  & 48" Samuel Oschin    & 1.012          & Mt. Palomar, California &              &     & S,R &       &              &       \\
\end{tabular}
\raggedright
\footnotesize{
R indicates repository for data retrieval. S indicates search capability.\\
NOIR: NSF National Optical Infrared Labs AstroArchive (\url{https://astroarchive.noirlab.edu}).\\
ESO: European Space Organization Archive (\url{https://archive.eso.org}).\\
IRSA: NASA/CalTech Infrared Science Archive (\url{https://irsa.ipac.caltech.edu}).\\
SMOKA: NAOJ Subaru-Mitaka-Okayama-Kiso Archive Science Archive (\url{https://smoka.nao.ac.jp}).\\
SSOIS: CADC Solar System Object Image Search (\url{https://www.cadc-ccda.hia-iha.nrc-cnrc.gc.ca/en/ssois/}).\\
STScI: Space Telescope Science Institute (\url{https://www.stsci.edu/}).}
\label{tab:equipAndArchives}
\end{table}

Table \ref{tab:equipAndArchives} lists the instruments and telescopes used in this work, along with their respective pixel scales, locations, and data archives.

\section{(248370) 2005 QN173 Data} 
\label{sec:ObjectData}

We provide current information regarding \objname{} below.

\begin{table}
    \centering
	\caption{\objname{} Properties}
	\begin{tabular}{lll}
		Parameter & Value & Source\\
		\hline\hline
		Discovery Date & 2005 August 29 & Minor Planet Center\\
		Discovery Observers & Near-Earth Asteroid Tracking (NEAT) & Minor Planet Center\\
		Discovery Observatory & Palomar & Minor Planet Center\\
		Activity Discovery Date & 2021 July 7 & CBET 4995\\
		Activity Discoverer(s) & A. Fitzsimmons / ATLAS & CBET 4995 \citep{fitzsimmons2483702005QN1732021}\\
		Orbit Type & Outer Main-belt & IMCCE, AstOrb\\
		Taxonomic Class & C-type (unconfirmed) & \citet{hsiehPhysicalCharacterizationMainBelt2021}\\
		Diameter & $D=3.6$~km; 3.4$\pm$0.4~km & Horizons, {\citet{harrisAsteroidsThermalInfrared2002}} ;  \citet{hsiehPhysicalCharacterizationMainBelt2021}\\ 
		Absolute $V$-band Magnitude & $H=16.02$ & Horizons\\
        Geometric Albedo & 0.054 & Horizons, \citet{mainzerNEOWISEDiametersAlbedos2019}\\
		Rotation Period & unknown & \\
        Orbital Period & $P=5.37$ yr & Horizons \\
		Semimajor Axis & $a=3.075$ au & Horizons\\
		Eccentricity & $e=0.226$ & Horizons\\
		Inclination & $i=0.067\degr$ & Horizons\\
		Longitude of Ascending Node & $\Omega=174.28$ & Minor Planet Center\\
		Mean Anomaly & $M=8.79\degr$ & Minor Planet Center\\
		Argument of Perihelion & $\omega=146.09\degr$ & Horizons\\
		Perihelion Distance & $q=2.374$ au & Horizons\\
		Aphelion Distance & $Q=3.761$ au & Horizons\\
		Tisserand Parameter w.r.t. Jupiter & $T_J=3.192$ & AstOrb\\
	\end{tabular}
\end{table}



\end{document}